\documentclass[sigconf,10pt]{acmart}

\copyrightyear{2026}
\acmYear{2026}
\setcopyright{cc}
\setcctype{by}
\acmConference[DTwin '26]{ACM Workshop on Digital Twins over NextG Wireless Networks }{October 26--30, 2026}{Austin, TX, USA}
\acmBooktitle{ACM Workshop on Digital Twins over NextG Wireless Networks (DTwin '26), October 26--30, 2026, Austin, TX, USA}
\acmDOI{10.1145/3842434.3844958}
\acmISBN{979-8-4007-2957-7/2026/10}

\usepackage{booktabs}
\usepackage{multirow}
\usepackage{tabularx}
\usepackage{makecell}
\usepackage{multicol}
\usepackage{bm}
\usepackage{tikz}
\usepackage{algorithm}
\usepackage{algpseudocode}
\usepackage{amsmath,amsfonts}
\usepackage{graphicx}
\usepackage{textcomp}

\usetikzlibrary{positioning, shapes, arrows.meta}

\AtBeginDocument{%
  }

\begin{document}

\title{HalifaxDT: A Wireless Digital Twin from Open Geospatial Data}

\author{Alexis Delplace} \affiliation{ \institution{Faculty of Computer Science, Dalhousie University} \city{Halifax} \country{Canada}} \affiliation{
\institution{LISN, Université Paris-Saclay} \city{Gif-sur-Yvette} \country{France}} \email{alexis.delplace@dal.ca} \author{Xinyi Li} \affiliation{ \institution{Faculty of Computer Science, Dalhousie University} \city{Halifax} \country{Canada}} \email{xn394804@dal.ca} 
\author{Samer Lahoud} \affiliation{ \institution{Faculty of Computer Science, Dalhousie University} \city{Halifax} \country{Canada}} \email{sml@dal.ca}
\author{Nur Zincir-Heywood} \affiliation{ \institution{Faculty of Computer Science, Dalhousie University} \city{Halifax} \country{Canada}} \email{zincir@cs.dal.ca} 

\begin{abstract}
Wireless digital twins can support site-specific analysis, planning, and experimentation for future wireless networks. However, constructing them at city scale remains challenging when accurate 3D city models are unavailable. This paper presents HalifaxDT, a wireless digital twin of the Halifax Peninsula in Nova Scotia, Canada, constructed from heterogeneous open geospatial and spectrum data. HalifaxDT combines building footprints, LiDAR elevation products, building metadata, and spectrum licensing records through a workflow that reconciles multiple sources of building-height information while preserving the provenance of geometry decisions. The resulting terrain, buildings, and gateway metadata are integrated into Sionna RT for wireless simulation. 
We evaluate HalifaxDT through two use cases. The first compares its coverage predictions with a reference derived from field measurements and with an analytical baseline. The second uses the digital twin to predict the received signal under normal operating conditions and detect interference. The coverage results show that HalifaxDT better preserves the spatial structure of the measured radio map than the analytical baseline. The interference study shows that deviations from the predicted reference can reveal interference that is difficult to detect from received power alone. We also identify current fidelity limitations, including simplified material representation, missing vegetation, and the need for broader RF calibration and synchronization with live measurements.
\end{abstract}



\maketitle

\section{Introduction}
\label{sec:introduction}

Wireless digital twins are becoming important tools for site-specific simulation, deployment planning, synthetic data generation, and what-if analysis in future wireless networks. Their usefulness, however, depends on the availability of an environment model that is sufficiently accurate for propagation analysis. In dense urban areas, radio performance is shaped by local geometry, blockage, reflection, diffraction, terrain elevation, antenna placement, and material properties. A wireless digital twin therefore cannot be treated only as a network model; it must also represent the physical environment in which the network operates.

Prior work spans simulation platforms, calibrated propagation models, and city-scale scene instantiations. Sionna RT~\cite{sionnaRT} provides the
differentiable ray-tracing engine used here, while measurement-calibrated
digital twins show how field observations correct systematic RSSI offsets and
improve fidelity~\cite{sudhakaran2024wireless}. City-scale systems such as
BostonTwin~\cite{testolina24} demonstrate detailed urban geometry when such geometry is already available. HalifaxDT complements these efforts by combining open-data scene reconstruction with sparse LoRaWAN field-measurement evaluation.

However, many cities lack detailed 3D models that can be directly converted into wireless simulation scenes. HalifaxDT therefore focuses on traceable reconstruction from open or institutional layers, including building footprints, LiDAR elevation products, volunteered geographic information, municipal records, and spectrum-licence databases. This is especially important for building-height estimation, where elevation rasters, explicit height attributes, floor-count estimates, and external topographic records differ in coverage, temporal validity, spatial alignment, and height definition.

We present HalifaxDT, a ray-tracing-ready wireless digital twin of the Halifax Peninsula in Nova Scotia, Canada. HalifaxDT is constructed from publicly available geospatial and spectrum data rather than from an existing 3D city model. It integrates municipal building footprints, LiDAR-derived elevation products, OpenStreetMap metadata, municipal building attributes, national topographic building records, and spectrum-licence information, and exports the resulting scene in a format compatible with Sionna RT. The generated scene assets, processed inputs, and reconstruction scripts are made available online for inspection and reuse~\cite{halifaxdtRepo}.

The contributions are threefold. First, we present a reusable workflow for constructing ray-tracing-ready wireless digital twins from heterogeneous open geospatial data. Second, we introduce a building-height inference process that combines elevation products, building attributes, floor-count estimates, and external records while preserving provenance. Third, we instantiate the workflow as HalifaxDT and evaluate it through coverage mapping and interference detection, showing how an open-data wireless digital twin can support both propagation analysis and jammer detection without labeled jammer data.

\section{HalifaxDT Construction Methodology}
\label{sec:construction}

The construction of a city-scale digital twin for wireless simulation can be
organized as a reusable four-stage workflow: (i) identifying the geospatial data
needed to describe the environment, (ii) aligning heterogeneous sources in a
common spatial frame, (iii) converting the aligned layers into a
ray-tracing-ready 3D scene, and (iv) integrating the resulting scene with a
wireless simulator. HalifaxDT instantiates this workflow for the Halifax
Peninsula, but the same procedure can be adapted to other cities where similar
open or institutional geospatial sources are available.

\subsection{Geospatial Inputs and Source Discovery}
\label{subsec:geospatial_inputs}

The first step is to identify the data layers required for the intended wireless simulation. If a georeferenced 3D city model is already available, the
construction problem is mainly one of conversion, coordinate normalization, and
radio-material annotation. Otherwise, the scene must be reconstructed from 2D
and 2.5D geospatial sources.

In practice, the required inputs can be grouped into a few main categories: existing 3D models, building footprints, elevation products, and building metadata.
Table~\ref{tab:dt_source_checklist} summarizes the role of each layer, typical
sources, and the corresponding HalifaxDT data. The selected geospatial layers
are integrated by spatially aligning them and transforming them into a common
metric coordinate system for scene generation.

\begin{table}[!h]
\centering
\caption{Core geospatial inputs for constructing a city-scale wireless digital twin.}
\label{tab:dt_source_checklist}
\scriptsize
\begin{tabular}{p{0.24\columnwidth}p{0.38\columnwidth}p{0.30\columnwidth}}
\toprule
Input & Typical sources & HalifaxDT example \\
\midrule
3D city model &
CityGML, OBJ/PLY, existing city models &
Not available; reconstructed from layers \\

Building footprints &
OSM, cadastral data, municipal/provincial GIS &
HRM Buildings; Shapefile \\

DSM &
LiDAR, photogrammetry, satellite DSM &
HRM LiDAR DSM; GeoTIFF \\

DEM/DTM &
LiDAR, national elevation datasets, terrain APIs &
HRM LiDAR DEM; GeoTIFF \\

Building metadata &
OSM tags, municipal building layers, building registries, topographic databases &
OSM/Overpass; GeoJSON, HRM Buildings CSV, NSTDB; Shapefile/GPKG \\
\bottomrule
\end{tabular}
\end{table}

Two elevation layers are especially useful when explicit building heights are
missing. A Digital Surface Model (DSM) describes the visible surface, including
buildings, vegetation, and other above-ground objects. A Digital Elevation Model
(DEM), or Digital Terrain Model (DTM), describes the bare-earth terrain. When a
DSM and DEM are co-registered, their difference provides height above the local terrain, which can be used to estimate building heights. This DSM--DEM strategy can
be generalized to other urban digital twins when co-registered surface and
terrain elevation products are available. In practice, the main challenges are
the availability and resolution of the elevation layers, their temporal
consistency with the building footprints, and the quality of the spatial
alignment between datasets.

HalifaxDT uses building footprints from Halifax Regional Municipality (HRM) as
the base geometry, HRM LiDAR products for elevation, and OpenStreetMap (OSM),
the HRM Buildings CSV, and Nova Scotia Topographic Database (NSTDB) records as
complementary sources of building attributes and height evidence
\cite{hrmBuildings,hrmLidarDsm2018,hrmLidarDem2018,nstdbBuildings,openStreetMap,overpassAPI}.
OSM provides volunteered height, floor/level, material, and roof tags when
available. The HRM Buildings CSV provides municipal building attributes,
including explicit height values and floor-count metadata, and NSTDB provides
an external building-height reference. The coverage and complementarity of
these sources are analyzed below.

\subsection{Preprocessing and Height Inference}
\label{subsec:preprocessing_height}

After source discovery, the vector and raster layers are normalized into a
common metric frame. In HalifaxDT, this step crops the HRM building footprints,
OSM/NSTDB building records, and HRM DSM/DEM rasters to the Halifax Peninsula,
reprojects them to the same local coordinate system, and links each footprint
to the raster samples and metadata records used for height inference. This normalization is required because the sources do not share common geometries or identifiers: municipal footprints, OSM polygons, and NSTDB records may describe the same building using different outlines, coordinate systems, and attributes. The specific projection and spatial extent are implementation-dependent. The key requirement is to place all geometry and raster evidence in a single local coordinate frame before scene generation.

The most important preprocessing task in HalifaxDT is building-height inference.
For a building footprint $b$, the DSM-derived height candidate is computed from
the distribution of normalized DSM--DEM samples intersecting the footprint:
\begin{equation}
h^{\mathrm{DSM}}_b =
Q_{0.95}\!\left(
\left\{
z_{\mathrm{DSM}}(u)-z_{\mathrm{DEM}}(u)
\,\middle|\,
u \in b,\ u \ \text{valid}
\right\}
\right),
\end{equation}
where $Q_{0.95}$ denotes the 95th percentile of valid raster samples. This
approximates the upper roof surface while being less sensitive than the maximum
to isolated vegetation, noise, missing raster values, or single-pixel outliers.
In practice, this estimate must still be robust to footprint misalignment,
roof-shape variability, multi-volume buildings, and differences between height
definitions used by external datasets.

Each data source can provide one or more candidate height estimates. Candidate height estimates can be derived from explicit height fields; floor or level counts combined with building-type-dependent floor-height priors; roof-height metadata; external elevation records corrected for local terrain elevation; or metadata such as building type, feature class, footprint area, and construction type. In HalifaxDT, these
non-DSM candidates are obtained from OSM tags, the HRM Buildings CSV, and NSTDB
records. OSM and HRM contribute explicit-height candidates when available,
falling back to level-count-derived candidates when level counts are provided.

\begin{figure}[t]
\centering
\includegraphics[width=\columnwidth]{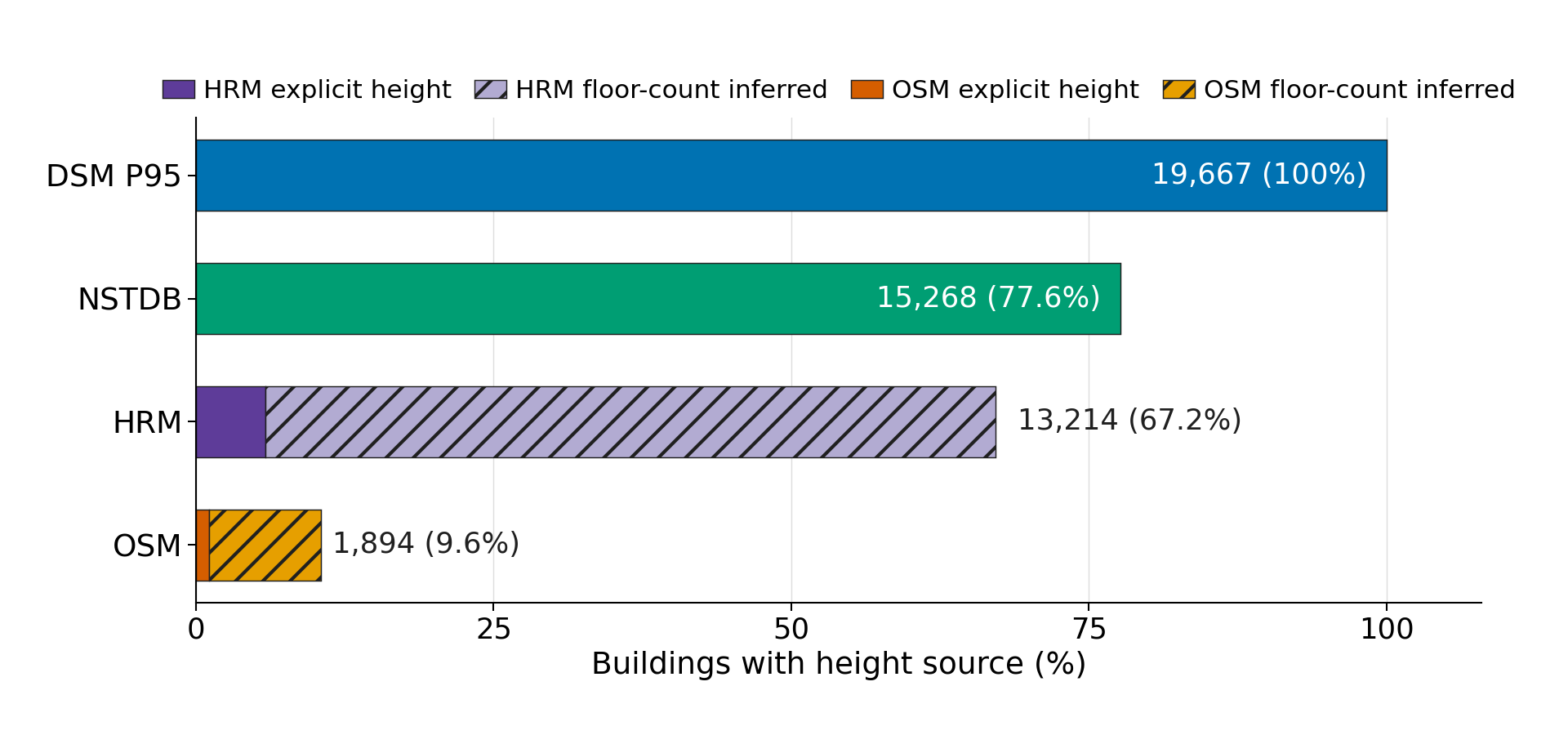}
\caption{Availability of building-height evidence across the sources integrated
in HalifaxDT.}
\label{fig:height_source_coverage}
\end{figure}

The source coverage observed in HalifaxDT motivates this multi-source design.
As shown in Figure~\ref{fig:height_source_coverage}, OSM provides
height-related metadata for only a limited fraction of buildings, with few
explicit heights. DSM-derived candidates are available for all footprints, while
NSTDB records and the HRM Buildings CSV add elevation-derived, explicit-height,
and floor-count-derived evidence. Because these sources differ in coverage,
derivation method, temporal validity, and spatial matching quality, combining
them reduces generic fallback heights and provides cross-checks where
independent candidates exist. Full DSM-derived candidate availability does not imply full reliable-height coverage: DSM-P95 may be affected by footprint alignment, roof shape, vegetation, temporal mismatch with the 2018 LiDAR acquisition, and height-definition differences. For high-elevation tails or mixed roof levels, the selector uses DSM-P75 rather than DSM-P95 to avoid letting small rooftop structures or adjacent objects determine the height of the whole building.

\begin{table}[t]
\centering
\caption{Final building-height source decisions in HalifaxDT.}
\label{tab:height_fusion_diagnostic}
\begin{tabular}{lcc}
\toprule
Selected height decision & Count & Share \\
\midrule
DSM-P95 & 16,163 & 82.2\% \\
DSM-P75 for high-tail profiles & 1,747 & 8.9\% \\
External or consensus source & 1,511 & 7.7\% \\
FCODE fallback & 246 & 1.3\% \\
\bottomrule
\end{tabular}
\end{table}

Figure~\ref{fig:height_source_coverage} reports source availability, whereas Table~\ref{tab:height_fusion_diagnostic} reports the resulting height-selection decisions. Most buildings retain DSM-derived heights, while multi-source evidence mainly affects multi-volume, externally confirmed, or invalid DSM cases. In 14.5\% of buildings, the final selected height differs from DSM-P95 by more than 2~m.

The height-inference protocol treats each possible height as a candidate
supported by one or more sources and assigns a deterministic rule-based
confidence score. Initial scores depend on source and derivation method, while adjustments use DSM pixel count, inner-footprint availability, roof-profile
stability, OSM/HRM overlap, OSM polygon ambiguity, NSTDB overlap or matching
distance, temporal consistency with the 2018 LiDAR acquisition date, and
physical plausibility. Scores only rank competing candidates for the same
building.

A DSM-derived value is informative when recent and well aligned, but less reliable for buildings constructed or modified after the DSM acquisition date.
This matters in HalifaxDT because the DSM and DEM come from 2018 LiDAR data and
the city has continued to grow. When construction or update metadata indicates
that a building postdates the DSM, the DSM-derived candidate is retained but
assigned lower confidence; buildings predating the DSM with good raster coverage
usually retain strong DSM evidence.

After scoring, the pipeline selects the final height. Physically plausible estimates that agree within a predefined tolerance are combined with a weighted median, which is robust to outliers while giving more influence to higher-scored candidates. The tolerance accounts for source-specific height definitions (e.g., roof base, maximum roof point, elevation-derived height, or level-derived height), so small discrepancies are expected. If no strong consensus exists, the highest-scored plausible candidate is selected; if direct evidence is missing or unreliable, the pipeline falls back to metadata-based inference.

The output of this stage includes both a height value and a provenance record:
the selected source or consensus, the decision strategy, the confidence level,
any post-DSM building flag, the candidates considered, and the reason for the
decision. This provenance supports manual inspection, later correction, and
adaptation of the same construction workflow to other cities with different
data availability.

\subsection{Scene Generation for Wireless Simulation}
\label{subsec:scene_construction}

The third stage converts the preprocessed geospatial layers into a
ray-tracing-ready scene. This requires choosing a terrain representation,
generating a ground mesh, extruding building footprints, assigning material
groups, and creating the scene description used by the ray tracer.

For terrain modeling, the main trade-off is between spatial resolution and the complexity of the resulting ground mesh. HalifaxDT uses the 5~m DEM for terrain
because terrain elevation varies more smoothly than building geometry at the
scale of the study area. A 1~m terrain surface would substantially increase mesh
complexity with limited benefit for the intended city-scale simulations. This
choice illustrates a general rule: high-resolution data should be preserved
where it affects propagation-relevant structure, but simplified where it mainly
increases computational cost.

Buildings are generated by extruding the height-enriched footprints. During
extrusion, building bases are sampled from the DEM so that walls follow local
terrain elevation. Footprint edges can be densified before extrusion to better
capture terrain variation along long building edges, and roofs are generated at
the selected roof elevation. These steps avoid treating buildings as flat
objects placed on a flat plane, while remaining simpler than full architectural
3D reconstruction.

The scene is also annotated with approximate radio-frequency materials.
Detailed facade materials are rarely available in open city-scale datasets. HalifaxDT therefore maps available semantic attributes to a small set of approximate RF material classes. HalifaxDT uses explicit OSM material tags when available and
otherwise infers coarse material labels from building type, feature class,
height, footprint geometry, and spatial priors. These labels are then mapped to
the material models available in the ray tracer. Terrain and common material
classes such as brick, concrete, metal, and wood can be assigned to their
corresponding or closest available RF material proxies.
Because city-scale material labels are sparse and coarse, these annotations are treated as optional scene metadata. The coverage-map evaluation below uses a single building material to avoid introducing an additional source of uncertainty from unvalidated material-label differences.

For larger cities, the same protocol can be applied per spatial tile while
preserving the mapping between local tile coordinates and the global coordinate
reference system. This makes the construction workflow scalable without changing
the underlying data model.

\subsection{HalifaxDT Instance and Sionna Integration}
\label{subsec:dataset_sionna}

The final stage packages the generated terrain, buildings, and RF material
annotations into a scene that can be loaded by Sionna RT.

The exported scene is organized around a Mitsuba-compatible XML scene file,
which is the scene format expected by Sionna RT. This XML file declares the
ITU radio-material definitions and references the mesh objects used by the
simulator. The terrain and building geometries are exported as PLY mesh files
and linked from the XML scene file, with material references assigned to each
mesh group. The scene description therefore acts as the central entry point
connecting the generated geometry with the RF material annotations.

Sionna RT can then load the XML scene and compute radio maps or propagation
paths using scenario-defined end-devices and gateways. In this way, the same
HalifaxDT scene can support multiple wireless analyses while keeping the
environment representation separate from experiment-specific radio
configuration.

\section{Use Cases}
\label{sec:use_cases}

\subsection{Coverage Mapping}
\label{subsec:coverage_mapping}

Coverage maps describe how radio-signal quality varies across a geographical
area. They support network planning, gateway placement,
coverage-hole identification, and the assessment of service availability before
or after deployment. Conventional coverage maps can be produced from field
measurements or analytical propagation models, but measurement campaigns are
costly and analytical models provide only a coarse representation of a specific
urban environment. A wireless digital twin provides an alternative for generating such maps: once a city and its radio infrastructure are represented in a
propagation-ready virtual environment, radio conditions can be estimated at
locations where no measurement is available and updated when the deployment
configuration changes.

\subsubsection{Coverage-Map Construction}

All maps use the same 25~m outdoor grid and geographical extent. The reference
map is derived from Long Range Wide Area Network (LoRaWAN) transmissions
collected on the Halifax Peninsula~\cite{delplace2025lrfhssCampaign}.

To construct a grid-level reference map, transmission attempts, including
received packets and non-receptions, are assigned to grid cells. Non-receptions
are treated as censored observations at $-140$~dBm, the minimum gateway
sensitivity observed in the dataset. Irregular measurements are then converted
into a continuous map using regression kriging~\cite{hengl2007regression},
which decomposes the estimated RSSI field into a large-scale trend and a kriged
residual:
\begin{equation}
\hat{r}(s) = \hat{m}(s) + \hat{\varepsilon}(s),
\end{equation}
where $\hat{m}(s)$ is the regression trend and $\hat{\varepsilon}(s)$ is the
kriged residual field. In HalifaxDT, the trend is modeled as
\begin{equation}
\hat{m}(s) =
\beta_0 + \beta_1 \log_{10}(d(s)) + \beta_2 d(s)
+ \beta_3 x(s) + \beta_4 y(s),
\end{equation}
where $d(s)$ is the distance to the gateway, and $x(s)$ and $y(s)$ are the
local projected coordinates of the measurement grid cell. The residual field is
estimated at each prediction location $s_0$ using local ordinary kriging:
\begin{equation}
\hat{\varepsilon}(s_0) =
\sum_{i=1}^{n} \lambda_i
\left(r(s_i)-\hat{m}(s_i)\right),
\end{equation}
where $r(s_i)-\hat{m}(s_i)$ is the residual at measurement grid cell $s_i$.
The weights $\lambda_i$ come from ordinary kriging with up to 18 neighboring
measurement cells within 300~m. A Gaussian smoothing step with standard
deviation 1.2 grid cells suppresses isolated interpolation artifacts. The resulting map is a measurement-derived interpolation product, not an independent ground-truth RSSI field; accordingly, we report measurement-support and kriging-uncertainty diagnostics over the evaluation mask. For the regression-kriging reference, the standard deviation over this mask has median,
90th percentile, 95th percentile, and maximum values of 5.51, 8.52, 9.10, and 10.27~dB, respectively.

The HalifaxDT coverage map is generated with Sionna RT on the HalifaxDT scene.
We use \texttt{PathSolver} point-by-point on the 25~m outdoor grid rather than
\texttt{RadioMapSolver}. For each valid grid point, the solver computes uplink
paths from a LoRaWAN end-device to the fixed gateway, relying on reciprocal path gain.
The end-device antenna is placed at terrain height plus 2.5~m to match the measurement setup, while the gateway antenna is 12~m above local ground level, corresponding to approximately 60~m above sea level over the Halifax Peninsula.
The simulation uses 904.6~MHz, 25~dBm end-device EIRP, 6~dBi gateway receive gain, isotropic vertically polarized antennas, \texttt{max\_depth}=16,
\texttt{samples\_per\_src}=20M, \texttt{max\_num\_paths\_per\_src}=500k, and
seed 42. Line-of-sight, diffuse reflection, and transmission/refraction are enabled; specular reflection and diffraction are disabled. Buildings use a single concrete material and terrain uses medium dry ground, keeping the comparison focused on geometry,
terrain, antenna placement, and ray-traced spatial structure rather than coarse material-label assumptions. The pointwise RSSI raster is floored at
\(-140\)~dBm and smoothed with a 3-by-3 median filter followed by Gaussian
smoothing with \(\sigma=62.5\)~m. This post-processing is applied only to
HalifaxDT because the analytical map is already spatially smooth. Simulations use \texttt{sionna-rt}~2.0.1 with the
\texttt{cuda\_ad\_mono\_polarized} Mitsuba variant.

For comparison, we retain the urban Okumura--Hata model, a well-established
analytical baseline for large urban environments around 900~MHz. It is evaluated
at every outdoor grid position using the same carrier frequency, fixed-gateway configuration, end-device height, antenna gains, and $-140$~dBm floor.

We first evaluate the maps without RSSI calibration. As reported in Table~\ref{tab:coverage_map_metrics}, HalifaxDT achieves lower MAE and better spatial metrics than Okumura--Hata. This raw comparison includes global link-budget mismatches, such as cable or connector losses, antenna-gain mismatch, and campaign-specific RSSI reporting effects that can penalize a map independently of its spatial structure, so we also evaluate a limited-data calibration setting.

\subsubsection{Limited-Data Calibration}

We consider a lightweight calibration scenario that corrects global RSSI offset
and scale errors from a small subset of the measurement campaign. Because these
records come from the same campaign used to construct the measurement-derived reference map, this is limited-data calibration rather than independent validation. To assess sampling variability, we repeat the procedure over 50 independent random draws; each draw selects thirty real transmission records,
including received RSSI values and non-receptions, to calibrate both models. We
use a left-censored Tobit model with latent RSSI,
\begin{equation}
 y_{i,m}^{*}=\alpha_m+\beta_m\widehat{r}_m(\mathbf{s}_i)+\varepsilon_i,
 \qquad \varepsilon_i\sim\mathcal{N}(0,\sigma_m^2),
\end{equation} 
where $\widehat{r}_m(\mathbf{s}_i)$ is the uncalibrated prediction of model $m$.
Parameter $\alpha_m$ represents a global offset, while $\beta_m$ adjusts the
RSSI scale. For received transmissions, the likelihood uses the observed RSSI.
For non-receptions, it uses the probability that the latent RSSI is below the
sensitivity floor $c=-140$~dBm:
\begin{equation}
 L_{i,m} =
 \begin{cases}
 \dfrac{1}{\sigma_m}\phi\!\left(
 \dfrac{r_i-\mu_{i,m}}{\sigma_m}
 \right), & i\in\mathcal{R},\\[1.2ex]
 \Phi\!\left(
 \dfrac{c-\mu_{i,m}}{\sigma_m}
 \right), & i\in\mathcal{C},
 \end{cases}
 \qquad
 \mu_{i,m}=\alpha_m+\beta_m\widehat{r}_m(\mathbf{s}_i),
\end{equation}
where $\mathcal{R}$ and $\mathcal{C}$ denote received and censored
non-received transmissions, respectively. This avoids treating every
non-reception as an exact RSSI measurement at the sensitivity floor.

With only 30 observations, an unconstrained affine calibration can overfit. We
therefore estimate \(\boldsymbol{\theta}_m=(\alpha_m,\beta_m,\sigma_m)\), with
\(\sigma_m>0\), by minimizing a ridge-regularized Tobit negative
log-likelihood,
\begin{equation}
 \widehat{\boldsymbol{\theta}}_m=
 \arg\min_{\boldsymbol{\theta}_m}
 \left[-\frac{1}{30}\sum_{i=1}^{30}\log L_{i,m}
 +\lambda_{\alpha}\left(\frac{\alpha_m}{10}\right)^2
 +\lambda_{\beta}(\beta_m-1)^2\right],
\end{equation}
with \(\lambda_{\alpha}=0.1\) and \(\lambda_{\beta}=1\). These penalties shrink
the offset toward zero and RSSI scale toward one, keeping calibration close to
the unmodified map unless supported by the data. The calibrated raster reports
the fitted latent RSSI mean after affine calibration, while the likelihood
accounts for left censoring at \(-140\)~dBm. The calibrated comparison therefore
emphasizes how well each map preserves the spatial organization of the
measurement-derived reference.

\subsubsection{Spatial Interpretation}

\begin{figure}[t]
\centering
\includegraphics[width=\columnwidth]{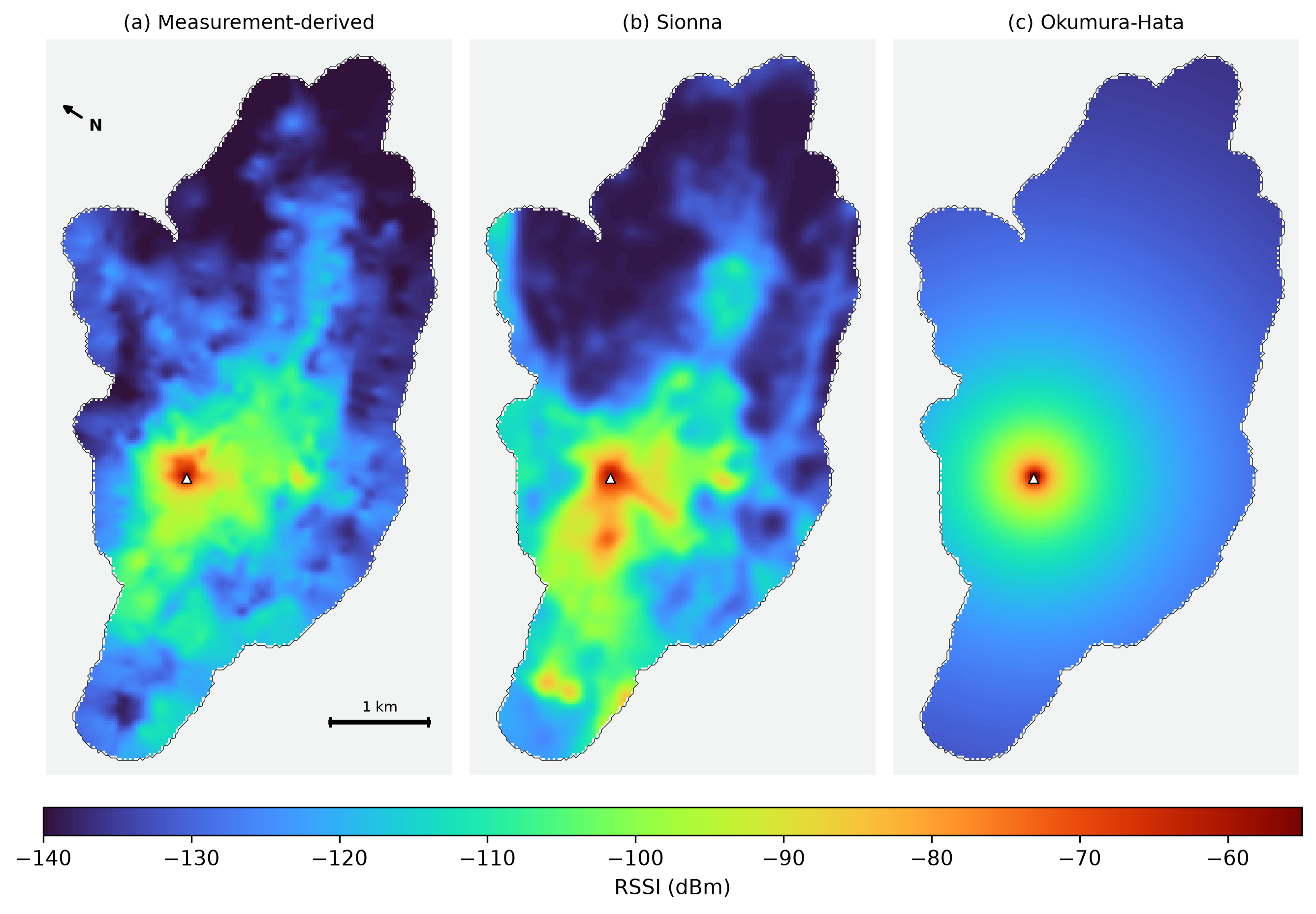}
\caption{Outdoor coverage map of Halifax: measurement-derived reference map
(left) and Tobit-calibrated maps from HalifaxDT (center) and Okumura--Hata
(right).}
\label{fig:coverage_map_comparison}
\end{figure}

Figure~\ref{fig:coverage_map_comparison} provides a spatial view of the
differences summarized in Table~\ref{tab:coverage_map_metrics}.
The Okumura--Hata map follows a smooth radial attenuation
trend but cannot reproduce site-specific shadowing and irregular coverage
boundaries. In contrast, the HalifaxDT map exhibits spatial variations aligned
with those visible in the measurement-derived map. We quantify this observation
using global and spatial-comparison metrics in
Table~\ref{tab:coverage_map_metrics}.

The comparison also shows how scene fidelity affects the radio map. In the real southern park area, dense vegetation partly obstructs propagation, whereas HalifaxDT represents this area as relatively flat and open. The model therefore allows stronger direct and reflected paths between end-devices and the gateway, leading to RSSI overestimation. Two localized building-adjacent overestimation areas likely reflect similar missing vegetation losses. These errors reflect environment-model limitations and show where
additional scene detail would most improve the digital twin.

\subsubsection{Quantitative Comparison}

\begin{table}[t]
\centering
\caption{Global and spatial coverage-map metrics. Arrows indicate the preferred
direction, and Gain denotes the relative improvement of HalifaxDT over
Okumura--Hata. For the 30-point calibration rows, each entry reports the central value over 50 independent random calibration draws as mean $\pm$ standard deviation.}
\label{tab:coverage_map_metrics}
\setlength{\tabcolsep}{3pt}
\renewcommand{\arraystretch}{1.08}
\scriptsize
\resizebox{\columnwidth}{!}{%
\begin{tabular}{@{}lllllll@{}}
\toprule
Cal. & Model
& MAE$\downarrow$
& V-RMSE$\downarrow$
& FSS$\uparrow$
& IoU$\uparrow$
& BDist$\downarrow$ \\
& & (dB) & (dB$^2$) & & & (m) \\
\midrule
Raw & HalifaxDT
& \textbf{10.92} & \textbf{11.24} & \textbf{0.840}
& \textbf{0.664} & \textbf{220.4} \\
Raw & Okumura--Hata
& 20.66 & 16.10 & 0.679 & 0.503 & 469.5 \\
\emph{Gain} & {}
& \emph{+47.2\%} & \emph{+30.2\%} & \emph{+23.8\%}
& \emph{+31.9\%} & \emph{+53.1\%} \\
\midrule
30-pt & HalifaxDT
& \textbf{5.58 $\pm$ 0.31}
& \textbf{4.18 $\pm$ 1.42}
& \textbf{0.876 $\pm$ 0.008}
& \textbf{0.705 $\pm$ 0.014}
& \textbf{152.5 $\pm$ 5.0} \\
30-pt & Okumura--Hata
& 6.61 $\pm$ 0.35
& 14.68 $\pm$ 0.17
& 0.802 $\pm$ 0.018
& 0.621 $\pm$ 0.021
& 339.2 $\pm$ 31.9 \\
\emph{Gain} & {}
& \emph{+15.6\%} & \emph{+71.5\%} & \emph{+9.2\%}
& \emph{+13.6\%} & \emph{+55.0\%} \\
\bottomrule
\end{tabular}%
}
\end{table}

All metrics in Table~\ref{tab:coverage_map_metrics} use the same outdoor
evaluation mask, which keeps grid cells within 300~m of at least one raw field-measurement record used to construct the reference map. For the evaluation subset, the mask contains 28,852 grid cells over 18.03~km$^2$, with 4,958 raw records and 2,903 occupied measurement cells, corresponding to 274.9 records/km$^2$, 161.0 occupied measurement cells/km$^2$, and direct support in 10.1\% of evaluation cells. Across all available campaign records, the same region contains 7,435 records, including 4,926 received packets and 2,509 non-receptions. The mask therefore limits comparison to locally supported areas, ensuring that metrics are not dominated by regions where the reference map relies mainly on interpolation. MAE measures global pixelwise RSSI
accuracy: HalifaxDT reduces raw MAE by 47.2\% relative to Okumura--Hata, while after 50 repeated 30-point Tobit calibrations it remains 15.6\% better and both
models reach similar city-scale average error. This motivates the complementary
spatial metrics below.

As a non-interpolated sanity check, we evaluate received-only errors at the raw measurement locations used for the map comparison. Using 3,445 received records, calibrated HalifaxDT obtains MAE/RMSE of 7.96/9.84~dB, while calibrated Okumura--Hata obtains 7.57/9.38~dB. Thus, HalifaxDT's advantage in Table~\ref{tab:coverage_map_metrics} should be interpreted as improved measurement-supported coverage-map structure, not uniformly better packet-level RSSI prediction. Okumura--Hata captures the dominant city-scale attenuation trend, but its smooth radial structure cannot reproduce local coverage boundaries and shadowed regions as well as HalifaxDT. Non-receptions are excluded from this MAE/RMSE calculation and handled through censored Tobit calibration.

To evaluate spatial organization beyond average RSSI error, we use four
complementary metrics. Variogram RMSE (V-RMSE) compares predicted and reference
empirical semivariograms over the same lag distances~\cite{malicke2022scikitgstat};
after calibration, HalifaxDT is 71.5\% lower. Fractions Skill Score
(FSS)~\cite{roberts2008fss} compares local covered-area fractions while tolerating small spatial displacements and is 9.2\% higher. Intersection over Union (IoU) measures covered-area overlap~\cite{rezatofighi2019giou} and is 13.6\% higher. Boundary distance compares predicted and reference coverage
contours through symmetric nearest-contour distance and is 55.0\% lower. These
metrics show that HalifaxDT better preserves the site-specific structure of the
measured radio map. Without HalifaxDT smoothing, the uncalibrated map still outperforms Okumura--Hata in MAE, FSS, IoU, and boundary distance (12.30~dB,
0.845, 0.633, and 238.5~m), but V-RMSE increases to 63.17~dB$^2$, confirming its sensitivity to cell-scale spatial variation.

Overall, HalifaxDT is markedly more accurate without RSSI calibration, showing
that the digital-twin map requires substantially less empirical adjustment than
the analytical baseline. Across repeated 30-observation calibrations,
regularized censored calibration improves both models, yet HalifaxDT retains a
consistent advantage across the spatial coverage metrics. The coverage-map use
case therefore shows that HalifaxDT can provide measurement-supported spatial
coverage predictions beyond what is captured by a calibrated analytical
baseline.

\subsection{Interference Detection}
\label{subsec:interference_detection}

HalifaxDT can also provide a model-based reference for interference detection. 

Using the modeled environment and ray-traced channels, HalifaxDT estimates the
signal that a sensing unit should receive from legitimate transmitters under
normal operating conditions. This reference can be compared with real-world measurements~\cite{schosser2024advancing}; deviations may indicate unexpected
interference from malfunctioning equipment, unauthorized transmitters, or
jammers.

\subsubsection{Detection Method}
We consider a scenario with multiple legitimate transmitters and a jammer. The received signal at the sensing unit is modeled as
\[
y_{\mathrm{PT}} = \sum_{k=1}^{K} H_k x_k + H_{\mathrm{jam}} x_{\mathrm{jam}} + n
\]
where $H_k$ is the channel frequency response (CFR) between the $k$-th
legitimate transmitter and the sensing unit, $x_k$ is the transmitted OFDM
signal, and $n$ is additive white Gaussian noise.
The digital twin predicts the interference-free received signal as $y_{\mathrm{DT}} = \sum_{k=1}^{K} H_k x_k + n$. Interference is detected by computing the residual power $P_\varepsilon = \|y_{\mathrm{PT}} - y_{\mathrm{DT}}\|^2$.

\begin{figure}[htbp]
    \centering
    \includegraphics[width=\linewidth]{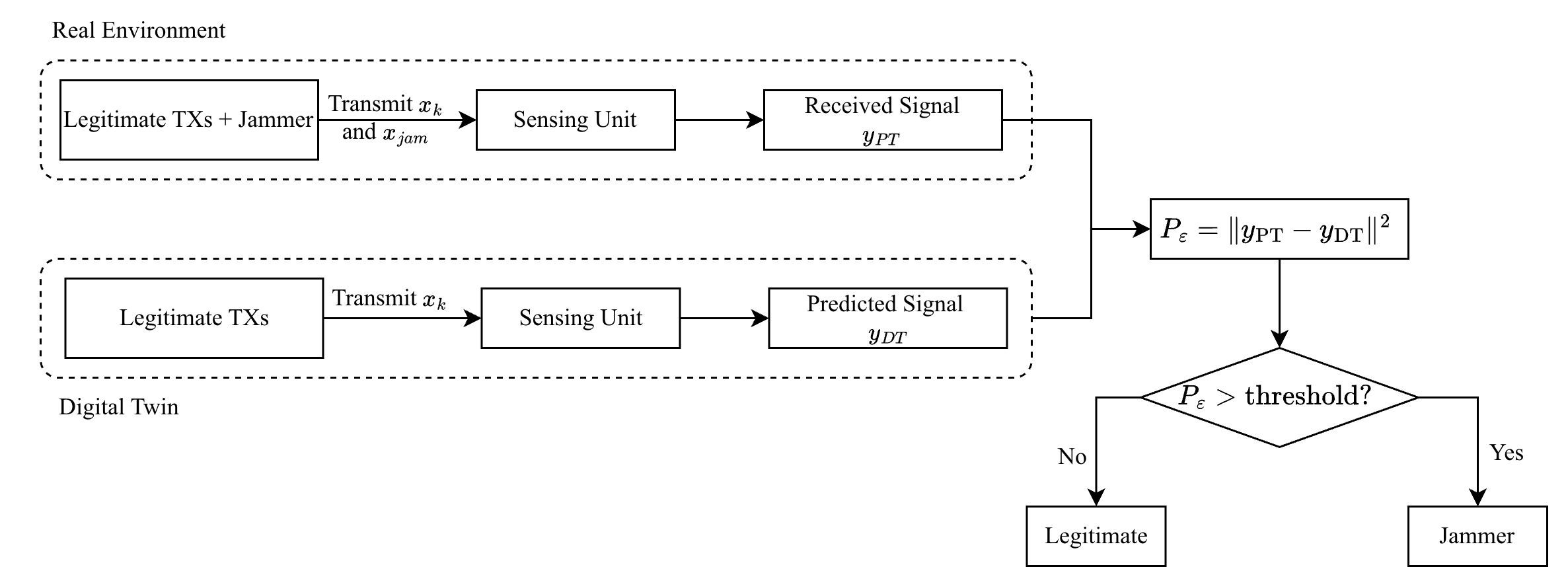}
    \caption{System overview of jammer detection using digital twin.}
    \label{fig:detection_overview}
\end{figure}

As shown in Fig.~\ref{fig:detection_overview}, the DT requires transmitter and sensing-unit locations, transmitted signals, and transmit powers to predict the interference-free signal. These inputs can be obtained through 5G positioning and network-side transmitter information, but may be noisy in practice. Here, they are assumed exact to isolate detection performance under an idealized DT.
A jammer is declared present when $P_\varepsilon$ exceeds a predefined
threshold.

We compare the DT-based detector with conventional energy detection (ED), which
uses received power $P_y = \| y_{\mathrm{PT}} \|^2$ without channel or
environment knowledge. Both methods are training-free and require no labeled
jammer data; legitimate samples are used only to calibrate the threshold.

\subsubsection{Experimental Setup}
\begin{figure}[htbp]
    \centering
    \includegraphics[width=0.8\linewidth]{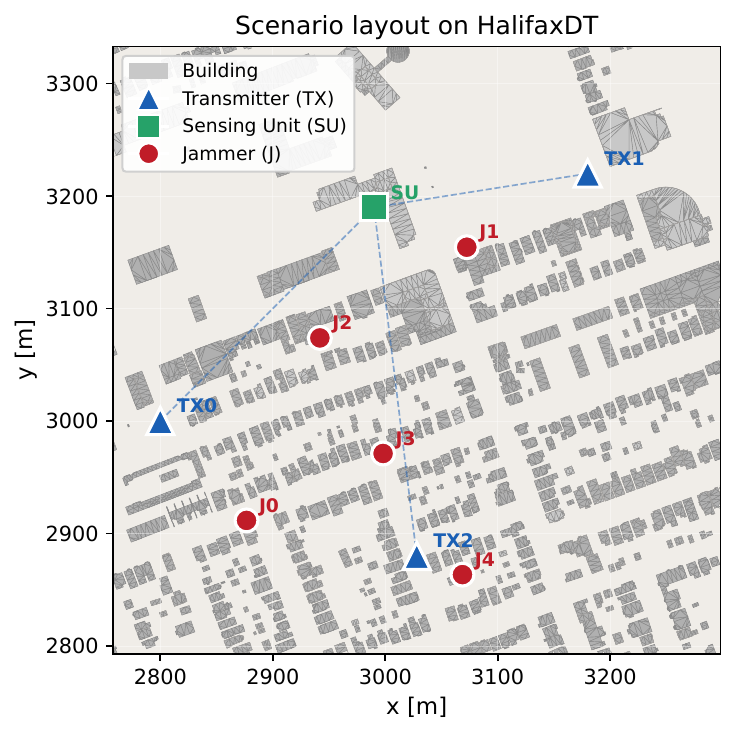}
    \caption{Simulation scenario layout.}
    \label{fig:detection_layout}
\end{figure}

We consider a $500 \times 500$ m subset of HalifaxDT with three fixed street-level
legitimate transmitters and one rooftop sensing unit at $(2990, 3190)$ m. Five
jammer locations are randomly sampled and evaluated independently, yielding five
test scenarios shown in Fig.~\ref{fig:detection_layout}.

The carrier frequency is $3.7$ GHz with $100$ MHz bandwidth. OFDM signals use $64$ subcarriers and $14$ symbols per frame, with $0$ dBm transmit power for both legitimate transmitters and the jammer. We consider eight jammer waveforms: continuous-wave jamming with amplitude variation (CWJ-A), continuous-wave jamming with frequency variation (CWJ-W), amplitude modulation jamming (AMJ), noise amplitude modulation jamming (NAMJ), narrowband noise jamming (NBNJ), multitone jamming (MTJ), linear frequency modulation jamming (LFMJ), and periodic pulse noise jamming (PPNJ), following the definitions in~\cite{ding2026fewshot}.

For both methods, the threshold is calibrated with $400$ legitimate samples to
target a $1\%$ false positive rate (FPR). At each jammer location, the test set
contains $500$ legitimate samples and $500$ jammer-contaminated samples for each
of the eight jammer types, i.e., $4{,}500$ samples per scenario: $500$
legitimate and $4{,}000$ jammer-contaminated.

The physical environment is simulated with the full HalifaxDT ray-tracing scene:
LoS and specular reflection are enabled, diffuse reflection and refraction are disabled, maximum interaction depth is 5, and a single vertically polarized isotropic antenna is used. The same scene generates both observations and the
DT reference, so this setting is an upper bound: $P_\varepsilon$ reflects detection under exact knowledge of
$H_k$ and does not account for DT-model mismatch. The DT reference
$y_{\mathrm{DT}}$ also includes an independently drawn noise term $n$, so the
residual reflects both jammer contribution and independent noise realizations.

\subsubsection{Results}
The detection performance is shown in Table~\ref{tab:detection_performance}.

\begin{table}[htbp]
    \centering
    \caption{Detection performance.}
    \label{tab:detection_performance}
        \begin{tabular}{crrrrr}
        \toprule
        Jammer & JNR (dB) & TPR$_{\mathrm{DT}}$ & TNR$_{\mathrm{DT}}$ & TPR$_{\mathrm{ED}}$ & TNR$_{\mathrm{ED}}$ \\
        \midrule
        J0 & $ 7.9$ & $99.8\%$ & $97.4\%$ & $76.0\%$ & $99.2\%$ \\
        J1 & $ 1.0$ & $100.0\%$ & $98.6\%$ & $70.8\%$ & $100.0\%$ \\
        J2 & $17.2$ & $100.0\%$ & $99.0\%$ & $100.0\%$ & $98.8\%$ \\
        J3 & $ 4.0$ & $100.0\%$ & $99.0\%$ & $95.2\%$ & $99.2\%$ \\
        J4 & $-8.1$ & $ 48.9\%$ & $98.8\%$ & $  4.5\%$ & $98.0\%$ \\
        \midrule
        Mean & -- & $89.7\%$ & $98.6\%$ & $69.3\%$ & $99.0\%$ \\
        \bottomrule
        \end{tabular}
\end{table}

The DT-based detector outperforms energy detection at all five jammer positions.
At the highest JNR (J2), both methods reach $100\%$ TPR because the jammer signal is strong. At moderate JNR
(J0, J1, J3), the DT-based method remains above $99\%$ TPR, while energy
detection ranges from $70.8\%$ to $95.2\%$. At the lowest JNR (J4), the
DT-based detector reaches 48.9\% TPR, compared with $4.5\%$ for energy detection. Its advantage comes from predicting the legitimate aggregate signal and testing the residual, whereas energy detection relies only on total received
power and cannot separate legitimate and interference components.

\section{Limitations and Future Work}
\label{sec:limitations_future_work}

HalifaxDT has several limitations that inform future work. Some scene inputs are
approximate or incomplete: the coverage-map evaluation uses a single concrete building material because city-scale material labels are sparse and coarse, and the 2018 LiDAR elevation data
may miss recently constructed or modified buildings. Vegetation is also omitted,
which can affect propagation accuracy, as observed in park areas. Finally, the current evaluation does not include independent RF measurement validation of the digital twin itself. In the interference-detection use case, the ground-truth received signal is also generated using the same ray-tracing scene, meaning the digital twin is assumed to represent the real environment. Validation against independent RF measurements and under DT-model mismatch is therefore an important direction for future work.

\section{Conclusion}
\label{sec:conclusion}
This paper presented HalifaxDT, a propagation-oriented wireless digital twin of
the Halifax Peninsula built from open geospatial and spectrum data. Its
reproducible workflow transforms heterogeneous 2D and elevation sources into a
ray-tracing-ready urban scene with inferred building heights, terrain elevation,
optional coarse material metadata, and gateway metadata. Through coverage
mapping and interference detection, HalifaxDT shows how an environment-aware
digital twin can provide spatial and signal-level information not captured by
coarse analytical propagation models or received-power-only detection. The
results also show that performance depends on scene fidelity: missing
vegetation, simplified material representation, and limited RF calibration can
create systematic errors, motivating measurement-driven calibration, richer
environmental modeling, and tighter synchronization with live wireless
observations.

\begin{acks}
The authors gratefully acknowledge Noah Jehanno and Nathan Chappe for their contributions to the data collection and prototyping activities that supported this work.
\end{acks}

\bibliographystyle{ACM-Reference-Format}
\bibliography{references}

\end{document}